\documentclass[sigconf]{acmart}

\usepackage{tikz}
\usepackage{graphicx}
\usepackage{tcolorbox}
\usepackage{multirow}
\usepackage{makecell}
\usepackage{threeparttable}
\usepackage{array}
\usepackage{listings}
\usepackage{tabularx}
\usepackage{xcolor}
\usepackage{fancyvrb} 
\usepackage{fontawesome} 
\usepackage{enumitem}
\usepackage{array}
\usepackage{booktabs} 
\usepackage{pifont}
\usepackage[font=small]{caption} 
\usepackage{titlesec}
\usepackage{balance}

\definecolor{c1}{HTML}{F6D4CF}

\newif\ifSimpleMode\SimpleModefalse

\newcommand{\dx}[1]{\textcolor{black}{{#1}}}
\newcommand{\before}[1]{\textcolor{c1}{{\ifSimpleMode\relax\else#1\fi}}}

\setcopyright{none}
\renewcommand\footnotetextcopyrightpermission[1]{}
\acmConference[]{}{}{}
\acmYear{}
\acmISBN{}
\acmDOI{}
\acmPrice{}

\AtBeginDocument{%
  }
\begin{document}
\balance

\SimpleModetrue

\title{An Empirical Study of Vulnerable Package Dependencies in LLM Repositories}
\author{Shuhan Liu}
\affiliation{%
  \institution{Zhejiang University}
  \city{Hangzhou}
  \state{Zhejiang}
  \country{China}}
\email{liushuhan@zju.edu.cn}

\author{Xing Hu}
\authornote{Corresponding author: Xing Hu}
\affiliation{%
  \institution{Zhejiang University}
  \city{Ningbo}
  \state{Zhejiang}
  \country{China}}
\email{xinghu@zju.edu.cn}

\author{Xin Xia}
\affiliation{%
  \institution{Zhejiang University}
  \city{Hangzhou}
  \state{Zhejiang}
  \country{China}}
\email{xin.xia@acm.org}

\author{David Lo}
\affiliation{%
  \institution{Singapore Management University}
  \country{Singapore}}
\email{davidlo@smu.edu.sg}

\author{Xiaohu Yang}
\affiliation{%
  \institution{Zhejiang University}
  \city{Hangzhou}
  \state{Zhejiang}
  \country{China}}
\email{yangxh@zju.edu.cn}

\begin{abstract}
Large language models (LLMs) have developed rapidly in recent years, revolutionizing various fields. Despite their widespread success, LLMs heavily rely on external code dependencies from package management systems, creating a complex and interconnected LLM dependency supply chain. Vulnerabilities in dependencies can expose LLMs to security risks. While existing research predominantly focuses on model-level security threats, vulnerabilities within the LLM dependency supply chain have been overlooked. To fill this gap, we conducted an empirical analysis of 52 open-source LLMs, examining their third-party dependencies and associated vulnerabilities. We then explored activities within the LLM repositories to understand how maintainers manage third-party vulnerabilities in practice. Finally, we compared third-party dependency vulnerabilities in the LLM ecosystem to those in the Python ecosystem. Our results show that half of the vulnerabilities in the LLM ecosystem remain undisclosed for more than 56.2 months, significantly longer than those in the Python ecosystem. Additionally, 75.8\% of LLMs include vulnerable dependencies in their configuration files. This study advances the understanding of LLM supply chain risks, provides insights for practitioners, and highlights potential directions for improving the security of the LLM supply chain.
\end{abstract}

\maketitle
\section{Introduction}
In recent years, the rapid growth of large language models (LLMs) has led to the development of numerous powerful models, such as ChatGPT~\cite{achiam2023gpt} and LLaMA~\cite{touvron2023llama}. 
These models, leveraging progress in neural network-based machine learning architectures, can generate natural language text with fluency and coherence comparable to human writing~\cite{brown2020language,chowdhery2023palm,bubeck2023sparks}.
They are applied across diverse fields, such as software engineering, finance, and education~\cite{yang2024ecosystem,chang2024survey,fan2023large,wu2023bloomberggpt,kasneci2023chatgpt}.
LLMs rely extensively on reusable external code, typically in the form of packages, which are distributed via package management systems like PyPI~\cite{pypi}. This extensive dependency on external packages accelerates the development of LLMs and establishes a complex web of interconnected dependencies, collectively referred to as the \textit{LLM dependency supply chain}. This supply chain includes a wide range of third-party dependencies, each providing essential functionality for LLM development and deployment, such as data processing, model optimization, and performance enhancement. Although these dependencies are helpful for rapid development of LLMs, they also introduce security risks~\cite{duan2020towards}. Vulnerabilities in any part of the supply chain can compromise the integrity and security of the model.

As LLMs gain popularity, researchers are increasingly focusing on their security. Current studies mainly address external threats such as adversarial attacks, data extraction risks, and the generation of harmful outputs~\cite{chen2023jailbreaker,carlini2021extracting, wan2023poisoning}, with efforts concentrated on defenses like inference guidance and output filtering~\cite{chiang2023vicuna,alon2023detecting}. Hu et al.~\cite{hu2024large} analyzed 59 articles from recent surveys and found that 19 focused on model-level security, while 24 only examined ChatGPT,  neglecting the broader supply chain. \before{Recent studies have explored components of the LLM supply chain~\cite{hu2024large, wang2024large}, providing a high-level overview of vulnerabilities across the supply chain, including infrastructure, the model lifecycle, and the application ecosystem. However, these studies lack an in-depth analysis of real-world vulnerabilities and fail to explore the LLM dependency supply chain.}
\dx{Recent studies have examined components of the LLM supply chain~\cite{hu2024large, wang2024large}, offering high-level overviews of security risks. Hu et al.\cite{hu2024large} introduce a taxonomy of 12 potential security risks, ranging from upstream data poisoning to downstream deployment vulnerabilities. Wang et al.\cite{wang2024large} discuss open challenges in infrastructure, lifecycle management, and threat modeling. However, these studies lack an in-depth analysis of real-world vulnerabilities and fail to explore the LLM dependency supply chain.}

\dx{To address this gap, we focus on security vulnerabilities in the LLM dependency supply chain. Specifically, we conduct an empirical analysis consisting of three key steps:
\ding{182} identifying third-party dependencies in 52 open-source LLM projects by parsing their source code and configuration files;
\ding{183} collecting vulnerability data from Snyk.io~\cite{snyk_security} and Libraries.io~\cite{libraries_io}, including CVEs, CWEs, severity levels, and lifecycle metadata; and
\ding{184} analyzing real-world patching practices by examining merged pull requests (PRs) related to vulnerabilities to understand how maintainers discuss and fix these issues in practice.}

In particular, we investigated the following three research questions:

\noindent \textbf{RQ1: What are the characteristics of the dependencies of LLM?}
RQ1 aims to examine the characteristics of LLM dependency packages. We analyze LLM dependencies in terms of usage patterns and domain distributions. Our findings reveal that most LLM dependency packages are sparsely utilized, with only 11\% appearing in more than 10 repositories across 52 LLMs, while 60.1\% are used only once. High-frequency dependencies are commonly sourced from popular repositories. 
In terms of domain distribution, LLM dependencies are predominantly focused on data processing, machine learning support, and system integration.

\noindent \textbf{RQ2: What are the vulnerabilities in the LLM dependency supply chain?}
In RQ2, we examine vulnerabilities in the LLM dependency supply chain, focusing on the vulnerabilities of packages used by LLMs. Our investigation covers trends, types, lifecycles of vulnerabilities, and their impact on LLMs.
We find that most vulnerabilities are classified as medium or high severity. Among packages with vulnerability reports, a median of 87.5\% of their versions are affected. Regarding vulnerability types, Denial of Service (DoS) is the most prevalent CWE in the LLM supply chain.  
Analyzing vulnerability lifecycles, we discover that half of the vulnerabilities remain undisclosed for more than 56.2 months, indicating substantial delays in their identification and disclosure.
Considering its influence on LLMs, we find that 75.8\% of LLMs with dependency configurations include at least one vulnerable dependency version. 

\noindent \textbf{RQ3: How are vulnerabilities in the LLM dependency supply chain addressed in practice?}
RQ3 investigates how maintainers handle dependency vulnerabilities in LLM projects. We explore vulnerability-related merged PRs in LLM repositories. Our analysis reveals that only 17.3\% of LLMs have vulnerability-related PRs, with 79\% of these PRs involving dependency updates. Additionally, 50\% of package version updates occur within 11 days of the release of a vulnerability-free package version.

To gain deeper insights into the characteristics of vulnerabilities in the LLM ecosystem, we compare vulnerabilities in the LLM dependency supply chain with those in the broader Python ecosystem. Our findings reveal that vulnerabilities in the LLM ecosystem tend to exhibit higher severity. Additionally, we observe that half of the vulnerabilities in the LLM ecosystem remain undiscovered for over 56.2 months, significantly longer than the 39 months observed in the Python ecosystem (see Section~\ref{sec:comparison}).

This paper makes the following contributions:

\begin{itemize}[leftmargin=*]

\item A comprehensive investigation of the dependency supply chain of 52 large language models, including the vulnerabilities in the supply chain. 
We update our comparison to the Python ecosystem by comparing our findings with a similar study on Python~\cite{alfadel2023empirical}.

\item A set of empirically grounded guidelines practitioners can adopt when implementing dependency management in their LLM projects. Our findings provide a foundation for improving the security and resilience of LLM ecosystems.

\end{itemize}

\section{Background}
\label{sec:backgroud}
In this section, we briefly review the relevant background.

\begin{figure}[h]
  \centering
  \includegraphics[width=\linewidth]{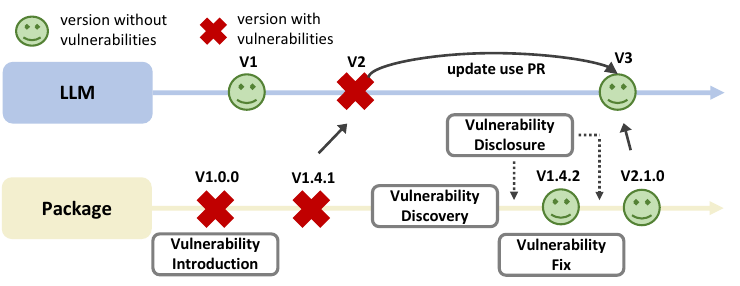}
  \caption{Vulnerability Lifecycle in LLM Dependency Supply Chain}
    \label{fig:vulnerabilityLifecycle}
\end{figure}

\before{\noindent \textbf{\textit{Large Language Model (LLM): }}The LLM refers to an artificial intelligence (AI) model that has been trained on large amounts of data and can generate text in a human-like fashion~\cite{brown2020language,fan2023large}. Some organizations publicly share the training code or model architecture of LLMs on open-source platforms (e.g., GitHub).}

\noindent \dx{\textbf{\textit{LLM repository: }}The LLM refers to an AI model that has been trained on large amounts of data and can generate text in a human-like fashion~\cite{brown2020language,fan2023large}. Some organizations publicly share the training code or model architecture of LLMs in repositories on open-source platforms (e.g., GitHub).}

\noindent \textbf{\textit{Software Dependency Supply Chain: }}The dependency supply chain refers to the network of software dependencies (e.g., external libraries) that a software project relies on~\cite{ohm2020backstabber}. This supply chain helps developers work faster by reusing code instead of starting from scratch. However, it also introduces risks as vulnerabilities in any component can propagate through the chain and affect dependent projects. As illustrated in Figure~\ref{fig:vulnerabilityLifecycle}, the vulnerability in package version V1.4.1 is introduced into the LLM when it adopts this package in version V2. The LLM subsequently addresses the issue by updating the dependency to the non-vulnerable package version V2.1.0.

\noindent \textbf{\textit{Vulnerability Lifecycle: }}The lifetime of a vulnerability typically involves several stages, from its initial introduction to discovery, fixing, and disclosure~\cite{rescorla2005security,arbaugh2000windows,frei2009security,kula2018developers}. As Figure~\ref{fig:vulnerabilityLifecycle} shows, we break the lifecycle into four main stages based on Alfadel et al.'s work~\cite{alfadel2023empirical}: \textit{\ding{182} Introduction:} The vulnerability is first introduced in the affected package, corresponding to the release date of the first affected version (e.g., V1.0.0 in Figure~\ref{fig:vulnerabilityLifecycle}). \textit{\ding{183} Discovery:} The vulnerability is identified through internal testing or external research. \textit{\ding{184} Fix:} The vulnerability is fixed in the package with the release of the first non-vulnerable version (e.g., V1.4.2 in Figure~\ref{fig:vulnerabilityLifecycle}). \textit{\ding{185} Disclosure:} The vulnerability is publicly disclosed on platforms like NVD or Synk. The vulnerability disclosure may happen before or after the fixed version (V1.4.2) is released.

\noindent \textbf{\textit{Snyk Vulnerability Database: }}Snyk~\cite{snyk_security} is a leading database for open-source vulnerabilities and cloud misconfiguration. It provides comprehensive information on vulnerabilities for specific packages, including details such as the CVE ID~\cite{cve}, CVSS score~\cite{cvss}, weakness type (CWE~\cite{cwe}), affected versions, and disclosure time. 

\noindent \textbf{\textit{Libraries.io: }}Libraries.io~\cite{libraries_io} is a service that aggregates public information on open-source packages from various sources on the Internet. It collects data from numerous package managers, including PyPI~\cite{pypi}, and serves as a centralized resource for tracking package metadata, version histories, and dependency relationships.

\section{Data Collection}
\label{sec:datacollection}
\begin{figure*}[h]
  \centering  \includegraphics[width=\linewidth]{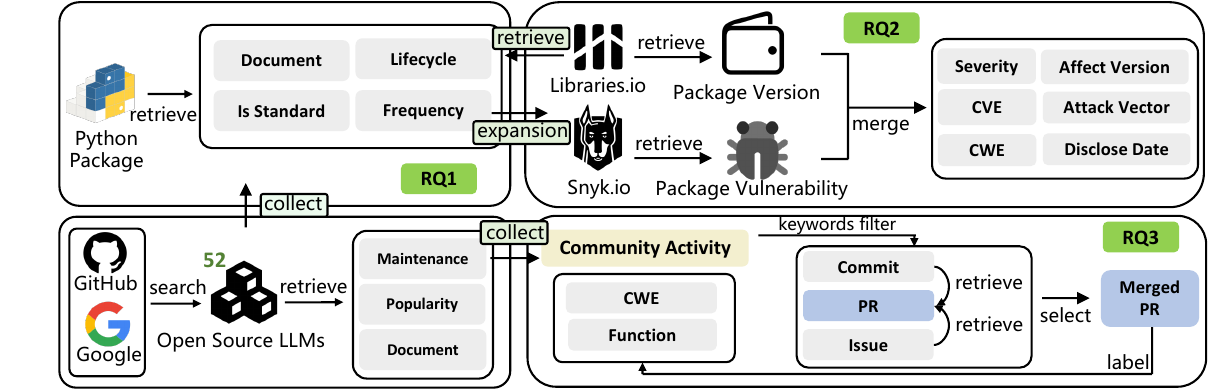}
  \caption{We provide an overview of our data collection process. First, we identify 52 open-source LLM repositories via GitHub and Google (bottom-left). We extract Python package dependencies from their codebases and retrieve related metadata from Libraries.io (RQ1). Next, we gather package version and vulnerability information from both Libraries.io and Snyk.io (RQ2). To investigate real-world vulnerability resolution practices, we collect community activities (commits, issues, and PRs), retaining only merged PRs for manual analysis (RQ3).}
    \label{fig:method}
\end{figure*}

\noindent In this section, we present our data collection procedure, as illustrated in Figure~\ref{fig:method}. To investigate supply chain vulnerabilities in LLMs, we collected open-source LLM projects from GitHub, identified their associated Python packages, and analyzed the vulnerabilities in these packages. We also examined community activities (e.g., issues, PRs, and commits) of LLM repositories. The resulting dataset comprises open-source LLMs, their related Python packages, identified supply chain vulnerabilities, and community activity records.

\dx{In this study, we consider both standard and non-standard Python libraries as part of the LLM dependency supply chain. While traditional software supply chain analyses primarily focus on external packages installed via package managers (e.g., PyPI), we include standard libraries for two key reasons. First, standard libraries (e.g., \texttt{os}, \texttt{json}) are widely used in LLM implementations and play critical roles in file access, system interaction, and data processing. Second, they are not immune to vulnerabilities: several historical CVEs have been reported in core Python modules. Although such vulnerabilities are less common, they can still propagate to downstream systems.}

\subsection{Open-Source LLMs}
\label{sec:opensourceLLMs}
To identify prevalent open-source LLMs, we followed the work of Hou et al.~\cite{hou2023large}, who reviewed 395 papers and identified 72 LLMs across three architectures: Encoder-Only, Encoder-Decoder, and Decoder-Only. Using their dataset as a foundation, we systematically searched GitHub for repositories corresponding to each LLM.
For LLMs without an identified repository, we conducted additional Google searches using keywords such as ``paper'', ``source code'', or ``repository''.
If this approach led to the discovery of a related academic paper, we examined the paper for references to replication links (e.g., a GitHub URL). 
When an LLM had multiple versions, each was included in the dataset and treated as a separate entity.
LLMs with no open-source repository identified through GitHub or Google searches were marked as unavailable.

\dx{To ensure comprehensive coverage of widely used open-source LLMs, we conducted additional searches using Google to identify popular LLMs that may not have been included in the systematic review~\cite{hou2023large}.
We adopted a multivocal literature review (MLR) approach~\cite{garousi2019guidelines}, which extends systematic literature reviews by incorporating both formal academic literature and grey literature (e.g., technical blogs, documentation, web pages, and community announcements).
These supplementary searches employed keywords such as ``large language models'' and ``open-source'' and focused on identifying widely publicized, community-supported models like ChatGLM~\cite{glm2024chatglm} and Qwen~\cite{bai2023qwen}.
This supplementary search led to the identification of seven open-source LLMs that were not captured during the initial screening.
In total, we built an initial dataset of 52 LLMs and their corresponding GitHub links.}

\before{To enhance coverage, we performed supplementary searches using Google to identify widely used models not included in Hou et al.’s review~\cite{hou2023large}. These searches focused on prominent, community-supported LLMs such as ChatGLM~\cite{glm2024chatglm} and Qwen~\cite{bai2023qwen}, leading to the identification of seven additional LLMs. This process resulted in a final dataset of 52 LLMs with corresponding GitHub links.}

\dx{Although HuggingFace hosts a wide range of pre-trained models, we did not systematically include them in our dataset. Most models on HuggingFace lack accessible source code repositories, which are essential for conducting code-level analysis of third-party dependencies. Therefore, we restricted our study to models hosted on GitHub, where full source code is typically available.}
For each LLM, we collected project-specific details, including:

\begin{itemize}[leftmargin=*]
\item \textbf{Documentation: }We gathered each project's \texttt{Readme}, \texttt{setup.py}, and \texttt{requirements.txt} files, which provided information about the project and the configuration of the Python packages.
\item \textbf{Popularity: }To evaluate the significance and activity of the projects, we employed the GitHub API to retrieve metrics including star counts and fork counts. These metrics provided valuable indicators of the popularity of the LLMs.
\item \textbf{Maintenance: }To elucidate the maintenance status of the projects, we collected metrics including issue counts, providing insights into maintenance activities and project engagement levels.
\end{itemize}
Table~\ref{tab:llmArchitecture} presents the distribution of LLM architectures and their associated package dependencies in our dataset.
\dx{Decoder-Only architectures constitute the majority (30 out of 52 models), followed by Encoder-Decoder and Encoder-Only architectures. On average, the LLM repositories have 5,188 GitHub stars, indicating strong community interest and relevance. Additionally, 58\% (30 out of 52) of the repositories include a dependency configuration file (i.e., \texttt{requirements.txt} and \texttt{setup.py}), specifying an average of 11.9 dependency entries per file. This suggests that a substantial portion of the collected LLM projects provide explicit and structured dependency configurations, which facilitates reliable analysis of their dependency supply chains.}
\begin{table}[h!]
  \caption{Overview of LLM Architectures and Package Dependencies}
  \label{tab:llmArchitecture}
  \footnotesize
  \resizebox{\linewidth}{!}{
  \begin{tabular}{@{}c|ccc|c@{}}
    \toprule
    \textbf{LLM Type} & \textbf{Decoder-Only} & \textbf{Encoder-Decoder} & \textbf{Encoder-Only} & \textbf{ALL} \\
    \midrule
    \textbf{LLM Num} & 30 & 12 & 10 & 52\\
    \textbf{Package Num} & 41 & 34 & 37 & 39 \\
    \textbf{Requirements} & 20 & 4 & 6 & 30 \\
    \scriptsize{\textbf{Requirements Package}} & 10.85 & 9.75 & 17.7 & 11.9 \\
    \textbf{Stars (median)} & 7,155 & 1,257 & 2,447 & 5,188\\
    \textbf{Created Year (avg)} & 2022 & 2020 & 2019 & 2021\\
    \bottomrule
  \end{tabular}}
\end{table}

\subsection{LLM-related Python Package}
\label{sec:LLMRelatedPythonPackage}
We then identified the Python packages used in the collected open-source LLMs (Section~\ref{sec:opensourceLLMs}) by parsing \textit{import} and \textit{from … import} statements in the repository files.
To classify a package as either part of Python’s standard library or an external dependency (typically installed via package managers like PyPI), we cross-referenced each identified package with the official Python Standard Library documentation~\cite{python_docs} and Libraries.io~\cite{libraries_io} for verification. These resources supplemented package information: Python documentation detailed the functionality of the standard package, while \texttt{Libraries.io} provided non-standard package metadata (e.g., descriptions, release history, and popularity metrics). Through this process, we identified 482 distinct packages used in LLM projects, including 94 standard packages and 388 non-standard packages. \dx{Different versions of the same package (e.g., \texttt{torch==1.10.0} and \texttt{torch==2.0.1}) are counted as a single entry.}

\subsection{Supply Chain Vulnerabilities}
We retrieved package version information from Libraries.io~\cite{libraries_io} and security vulnerability data from Snyk.io~\cite{snyk_io}.
Libraries.io provides metadata for PyPI packages, which includes details such as available package versions and their creation timestamps~\cite{alfadel2023empirical}.
These details were crucial for mapping the affected versions identified in our vulnerability dataset and conducting time-based analyses. 
To collect vulnerability data, we relied on the dataset provided by Snyk.io, a platform that aggregates security reports across various package ecosystems, including PyPI. Snyk.io’s dataset includes detailed information for each vulnerability, such as its CVE, CWE, affected version ranges, and fixed versions. It also specifies the discovery and publication dates of vulnerabilities, enabling further temporal analysis. Vulnerabilities are categorized by severity based on their CVSS scores, ranging from critical, high, medium, to low.
In total, we collected 890 LLM dependency supply chain vulnerabilities and their related information.

\subsection{Community Activity}
\label{sec:communityActivity}
To analyze vulnerabilities discussed or identified during LLM project maintenance, we collected community activities, including commits, PRs, and issues. For automated identification, we applied the methodology proposed by Lai et al.\cite{lai2024securityweakness}, which employed regular expression rules designed to detect vulnerability-related terms in deep learning systems\cite{10.1145/2635868.2635880,10.1145/3106237.3117771}. These regex rules targeted keywords such as ``security'', ``vulnerability'', ``attack'', and ``CVE'' (refer to the replication package for the complete list). Using these rules, we systematically extracted vulnerability-related community activities.

\before{To ensure reliability, we focused on merged PRs, as they represent reviewed and accepted security fixes.}
We linked vulnerability-related commits and issues to their corresponding PRs, merging those retrieved from related commits and issues with explicitly vulnerability-related PRs. Only merged PRs were retained to capture confirmed resolutions rather than speculative or abandoned fixes, resulting in a dataset of 914 merged vulnerability-related PRs.

For specific vulnerability identification, we conducted a manual labeling process leveraging the authors’ development expertise. Two authors independently reviewed the merged vulnerability-related PRs, examining source code, titles, descriptions, comments, and discussions. Following the review, they labeled records across seven aspects: (1) CWE ID, (2) CWE Name, (3) CWE Pillar, (4) Root Cause, (5) Fixing Pattern, (6) Symptom and whether it is a (7) Vulnerability.
\dx{The CWE ID and CWE name follow the official CWE List~\cite{cwelist}, while the remaining four dimensions are designed to offer a more fine-grained understanding of real-world fixing practices and ensure structural consistency with supply chain vulnerabilities collected from Snyk~\cite{snyk_security}.}
The initial labeling yielded a near-perfect Kappa coefficient of 0.859~\cite{banerjee1999beyond}. Disagreements were resolved through discussions, leading to a refined classification strategy. Both authors then re-reviewed all data using the updated strategy, conducting comparative discussions to resolve disagreements across all seven aspects. As a result, we identified 100 real LLM vulnerabilities that were discussed and fixed within LLM community activities.

\section{RQ1: Dependency Feature}
\label{sec:RQ1}
\dx{This section addresses the research question: What are the characteristics of LLM dependencies?}

\subsection{Methodology}
To gain a comprehensive understanding of LLM dependencies, we analyzed the functionality of packages collected in Section~\ref{sec:LLMRelatedPythonPackage}.
We applied thematic analysis to package descriptions to analyze the themes. The process followed these steps~\cite{cruzes2011recommended}: (1) initial reading of the descriptions, (2) generation of initial codes for each package description, (3) identification of themes among the proposed codes, (4) review of the themes to identify merging opportunities, and (5) finalization of the themes. Steps (1) to (4) were performed independently by the first two authors. Following this, a series of meetings were held to resolve conflicts and assign the final themes (step 5). Finally, we identified seven main categories and 16 subcategories, assigning each package to a category.

\subsection{Results}
\subsubsection{Characteristics}
Across 52 LLMs, we identified 482 unique Python packages, comprising 94 standard library packages and 388 non-standard library packages. \dx{The large number of non-standard packages underscores the extensive reliance on third-party libraries to support functionalities beyond those offered by the Python standard library.}

\begin{figure}[h]
  \centering
  \includegraphics[width=\linewidth]{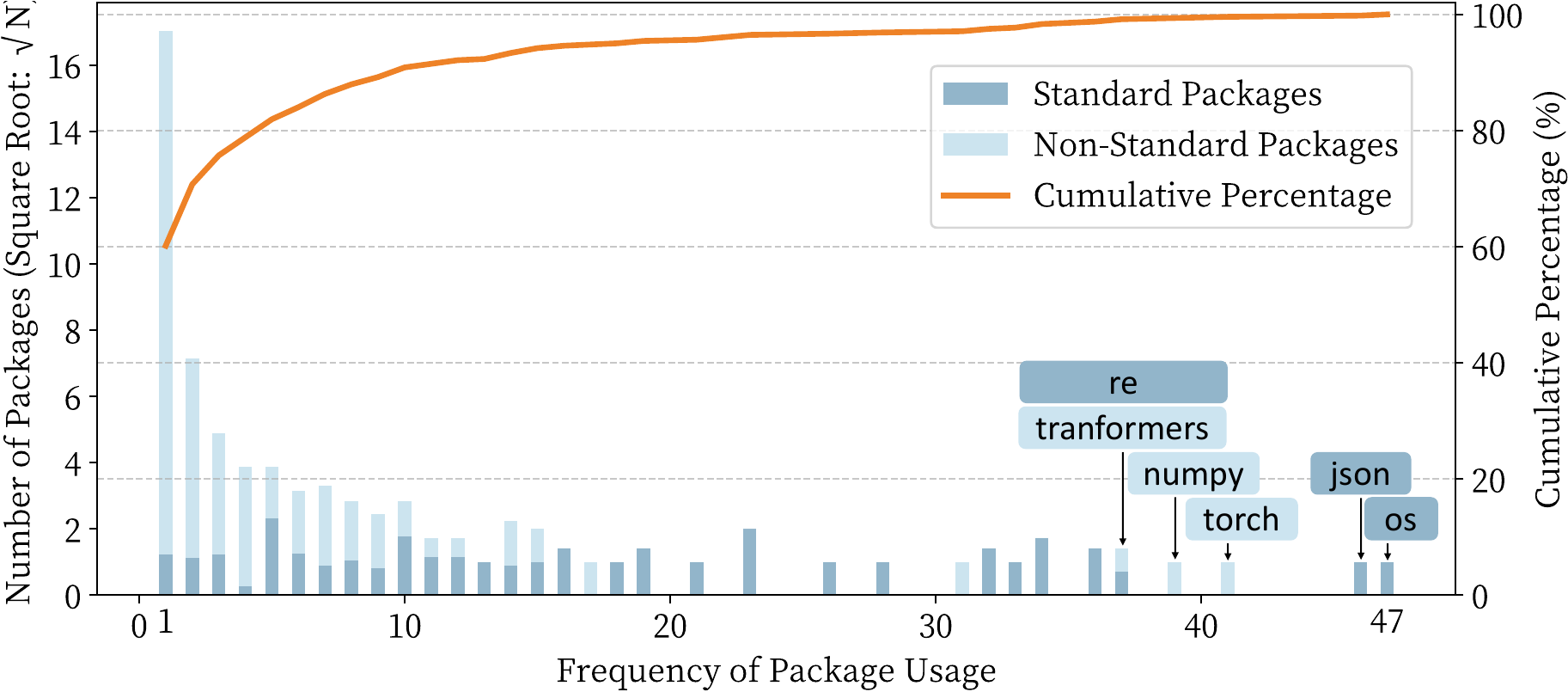}
  \caption{Package Usage Frequency with Cumulative Percentage}
    \label{fig:packagefrequency}
\end{figure}

\begin{figure}[h]
  \centering
  \includegraphics[width=\linewidth]{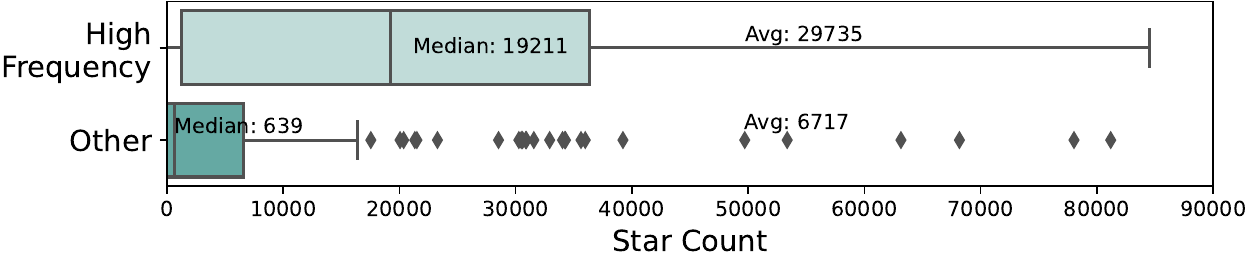}
  \caption{Comparison of Star Counts between High Frequency Packages and Other Packages}
    \label{fig:highFrequencyStar}
\end{figure}

Figure~\ref{fig:packagefrequency} illustrates the distribution of package usage frequencies across the analyzed LLMs. The data reveals disparities: 60.1\% of packages are used only once, while only 11\% appear more than 10 times, highlighting the prevalence of infrequently used dependencies.
To quantify this inequality, we calculated the Gini coefficient~\cite{gini1912variability} for package usage frequencies, which was 0.63. \dx{The Gini coefficient ranges from 0 (perfect equality) to 1 (perfect inequality), with values above 0.6 typically indicating a high degree of concentration or distributional imbalance.} Therefore, this high value underscores the reliance on a small subset of core dependencies that are disproportionately utilized, such as \texttt{json}, \texttt{os}, and \texttt{torch}, each referenced over 40 times. These high-frequency packages are present in more than 75\% of LLMs, playing a central role in enabling essential functionalities.
Further analysis reveals distinct patterns between standard and non-standard packages. We defined packages used 10 or more times across the 52 LLMs as high-frequency packages. \dx{This threshold was empirically determined based on the distribution in Figure~\ref{fig:packagefrequency}. As illustrated, the cumulative percentage curve flattens beyond a usage frequency of 10, indicating a natural divide between widely reused core packages and infrequently used ones.} High-frequency packages are evenly split between standard and non-standard categories, indicating the foundational role of standard libraries alongside specialized functionalities offered by non-standard packages. However, less frequently used dependencies are predominantly non-standard. While this fosters dependency diversity, it also introduces supply chain risks due to inconsistent usage and maintenance disparities.
\dx{For non-standard packages, we further examined differences between high-frequency and other packages. As shown in Figure~\ref{fig:highFrequencyStar}, high-frequency packages had a median star count of 19,211, indicating substantial popularity. To assess the normality of the star count distributions, we conducted Shapiro–Wilk tests~\cite{shapiro1965analysis}. The results (W = 0.7798, p = 0.0021 for high-frequency; W = 0.4926, p \textless ~0.0001 for low-frequency) indicate significant deviation from normality in both groups. Consequently, we applied the Mann–Whitney U test~\cite{mann1947test}, which revealed a statistically significant difference (U = 3,111, p = 0.0043).}
\before{As shown in Figure~\ref{fig:highFrequencyStar}, the median star count for high-frequency packages was 19,211, reflecting their widespread popularity.}
\before{A Mann-Whitney U test~\cite{mann1947test} confirmed the statistical significance of this disparity (U = 9,762, p = 0.0001).}
Additionally, Cliff’s delta~\cite{cliff1993dominance} yielded a value of 0.3171, indicating a moderate effect.
These findings indicate that LLM package references are biased toward highly popular packages. However, this dependence on popular packages also introduces risks. The extensive use of these high-frequency packages means that any vulnerabilities within them could propagate across multiple LLM projects, amplifying the impact of potential security flaws.

\begin{center}
    \resizebox{\linewidth}{!}{
\begin{tabular}{l!{\vrule width 1pt}p{0.9\columnwidth}}
    \makecell{{\LARGE \faLightbulbO}}  &\textbf{Summary.} \textit{Only 11\% of packages in the LLM supply chain appear more than 10 times across 52 LLMs, and these packages are typically sourced from more popular repositories. 
    }\\
\end{tabular}}
\end{center}
\subsubsection{Domain Distribution} 
\label{sec:domainDistribution}
\begin{figure}[h]
  \centering
  \includegraphics[width=.7\linewidth]{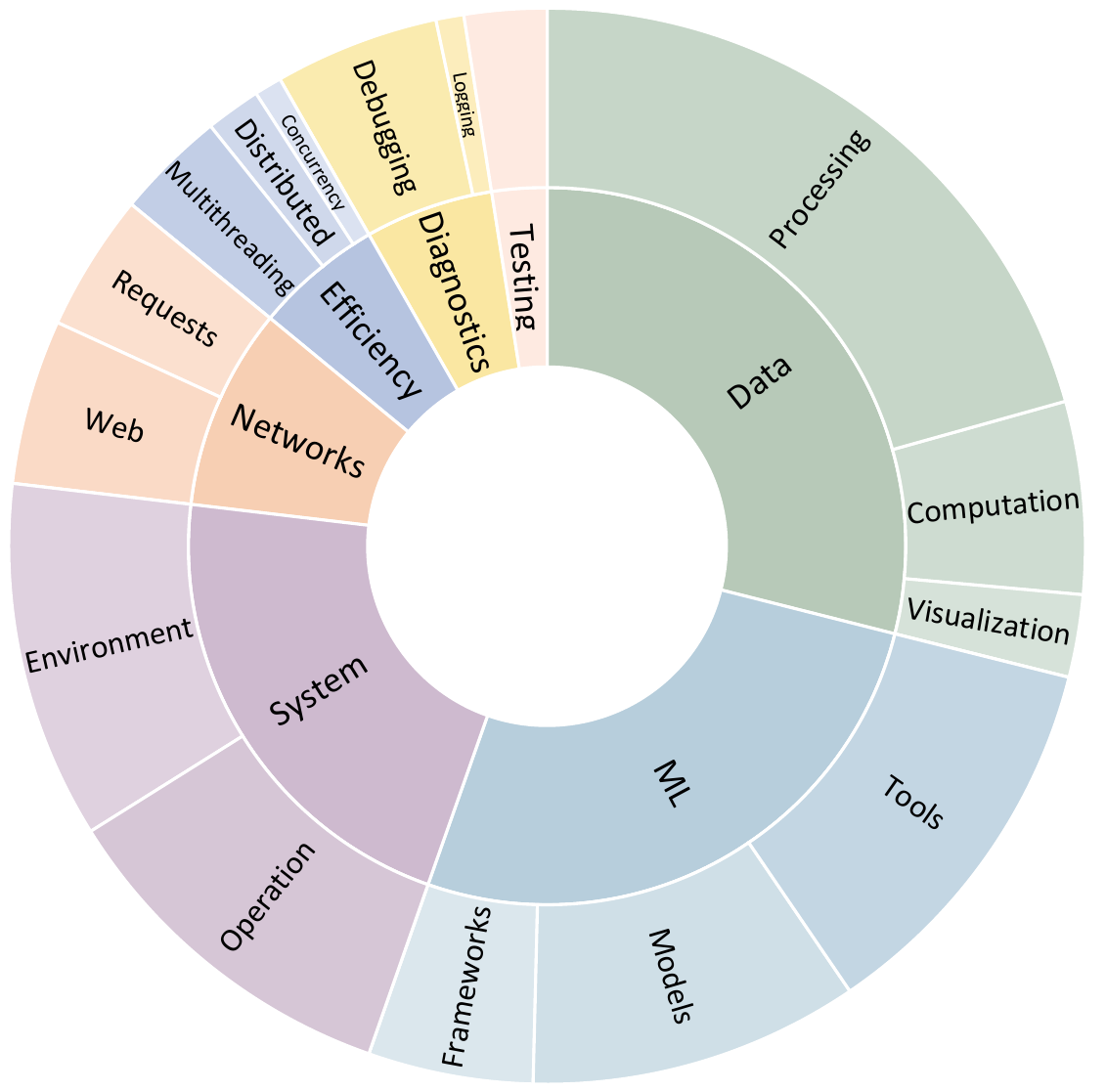}
  \caption{Functional Distribution of LLM-Related Packages}
    \label{fig:domainDistribution}
\end{figure}

Figure~\ref{fig:domainDistribution} shows the functional distribution of packages related to LLMs.

The most prevalent package types include ``Data Processing'' (25 packages), ``Machine Learning Tools'' (14 packages), and ``System Environment'' (13 packages), highlighting the ecosystem’s emphasis on data manipulation, development tools, and system integration. Additionally, ``Machine Learning Models'' and ``Machine Learning Frameworks'' collectively account for 18 packages, underscoring the critical role of foundational machine learning libraries in supporting LLM functionalities.
In contrast, specialized categories such as ``Distributed Efficiency'' (2 packages), ``Concurrency Efficiency'' (1 package), and ``Diagnostics Logging'' (1 package) are less frequent, suggesting these functionalities are less central to LLM development. Packages related to ``Networks'' (Web and Requests, totaling 11 packages) and ``Visualization'' (3 packages) emphasize the importance of external connectivity and data representation in the ecosystem.
Overall, this distribution reflects the versatility of LLM packages, balancing core machine learning support with specialized tools to enable robust development and deployment.

\begin{figure}[h]
  \centering
  \includegraphics[width=\linewidth]{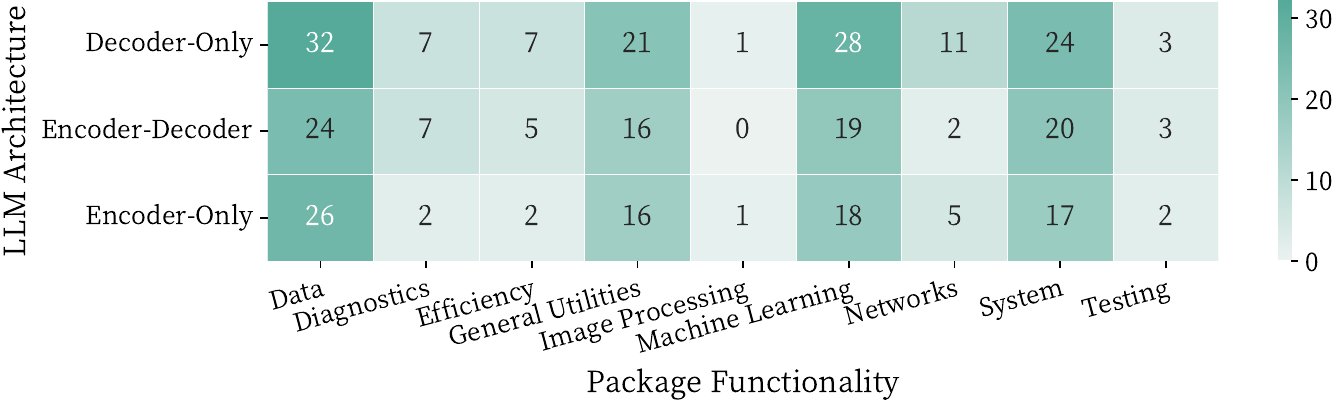}
  \caption{Package Functional Distribution across LLM Architectures}
    \label{fig:domainDistributionHeatmap}
\end{figure}

Figure~\ref{fig:domainDistributionHeatmap} illustrates the functional distribution of packages across different LLM architectures. Decoder-Only models employ a broader range of ``Networks'' packages than other architectures. This is due to their focus on real-time generative tasks, such as conversational AI. These tasks require robust networking capabilities for seamless data exchange and task-specific adaptations. For example, Decoder-Only models designed for real-time conversational AI often use networking packages like \texttt{requests} and \texttt{socket} to interact with external APIs, enabling dynamic and interactive responses during live interactions.
In contrast, Encoder-Decoder models show minimal reliance on ``Networks'' packages, reflecting their focus on offline tasks, such as machine translation or summarization. Encoder-Only models show limited usage of ``Diagnostics'' and ``Efficiency'' packages, suggesting a self-contained design that minimizes the need for extensive debugging or performance optimization tools. This is consistent with their primary application in understanding and classification tasks, where accuracy and interpretability are prioritized over interactivity or computational efficiency.
These differences underscore how architectural design influences package dependencies, with Decoder-Only models prioritizing interactivity, Encoder-Decoder models balancing task-specific functionality, and Encoder-Only models maintaining simplicity and independence.

\begin{center}
    \resizebox{\linewidth}{!}{
\begin{tabular}{l!{\vrule width 1pt}p{0.9\columnwidth}}
    \makecell{{\LARGE \faLightbulbO}}  &\textbf{Summary.} \textit{LLM-related packages primarily support data processing, machine learning, and system integration. The architectural differences among LLMs influence the functional roles of their dependencies.}\\
\end{tabular}}
\end{center}

\section{RQ2: Vulnerabilities}
\label{sec:RQ2}
\dx{This section aims to answer the research question: What are the vulnerabilities in the LLM dependency supply chain?}

\subsection{Methodology}
\label{sec:RQ2Method}

To examine vulnerability trends in the LLM dependency supply chain, we analyzed the temporal distribution of discovered vulnerabilities. Vulnerabilities were grouped by disclosure year to track their evolution alongside affected packages. We then categorized them by severity to identify trends across different risk levels. We also examined package versions impacted by vulnerabilities. In our analysis, any version containing at least one vulnerability was classified as vulnerable.

Each vulnerability was classified based on its CWE~\cite{cwe} category, as provided by Snyk.io, to facilitate the analysis of vulnerability types. For each CWE category, we assessed the number of vulnerabilities across different severity levels. This analysis enabled us to identify the most prevalent vulnerability types and their severity distribution. Additionally, we examined the distribution of vulnerabilities across packages, categorized by their functional roles as outlined in Section~\ref{sec:domainDistribution}. By mapping vulnerability types to package functions, we sought to determine whether certain vulnerability types are more common in packages with specific functions.

To analyze the lifecycle of vulnerabilities in the LLM dependency supply chain, we examined three key timestamps: (1) introduction date, marking the release date of the earliest package version affected by the vulnerability; (2) disclosure date, when the vulnerability was publicly reported on Snyk.io; and (3) fix date, indicating the release of the patched package version.
We first measured the time between introduction and disclosure to assess disclosure delays. 
\dx{Kaplan-Meier} survival analysis (also known as event history analysis)~\cite{Kaplan01061958} was employed to model the likelihood of undisclosed vulnerabilities over time. This method, widely used in prior research\cite{8721084,10.1145/3196398.3196401,alfadel2023empirical}, helps evaluate time-dependent risks.
Subsequently, we classified vulnerabilities into three categories based on their fix status: Fixed After Disclosure, Fixed Before Disclosure, and Never Fixed. This classification was determined using the vulnerability fix date and disclosure date as reference points. To analyze the relationship between severity and fix status, we calculated the number of vulnerabilities in each category across different severity levels.

To explore the impact of vulnerabilities in the LLM supply chain on LLMs, we focused on whether LLMs rely on vulnerable versions of packages. Specifically, we categorized each package version as ``vulnerable'' or ``not vulnerable'' based on the set of affected versions for each vulnerability.
Our analysis targeted LLMs with dependency configuration files (e.g., \texttt{requirements.txt} or \texttt{setup.py}) to identify whether the specified package versions were vulnerable. 
We considered both strict version constraints (e.g., ==) and loose version constraints (e.g., \textgreater=). For loosely constrained dependencies, we evaluated whether the specified version range included any vulnerable versions.

\subsection{Results}
\subsubsection{Vulnerability Trends}
\label{sec:vulnerabilityTrends}
We analyzed temporal trends in vulnerabilities, their distribution by severity, and their impact across package versions in the LLM supply chain.

\begin{figure}[h]
  \centering
  \includegraphics[width=\linewidth]{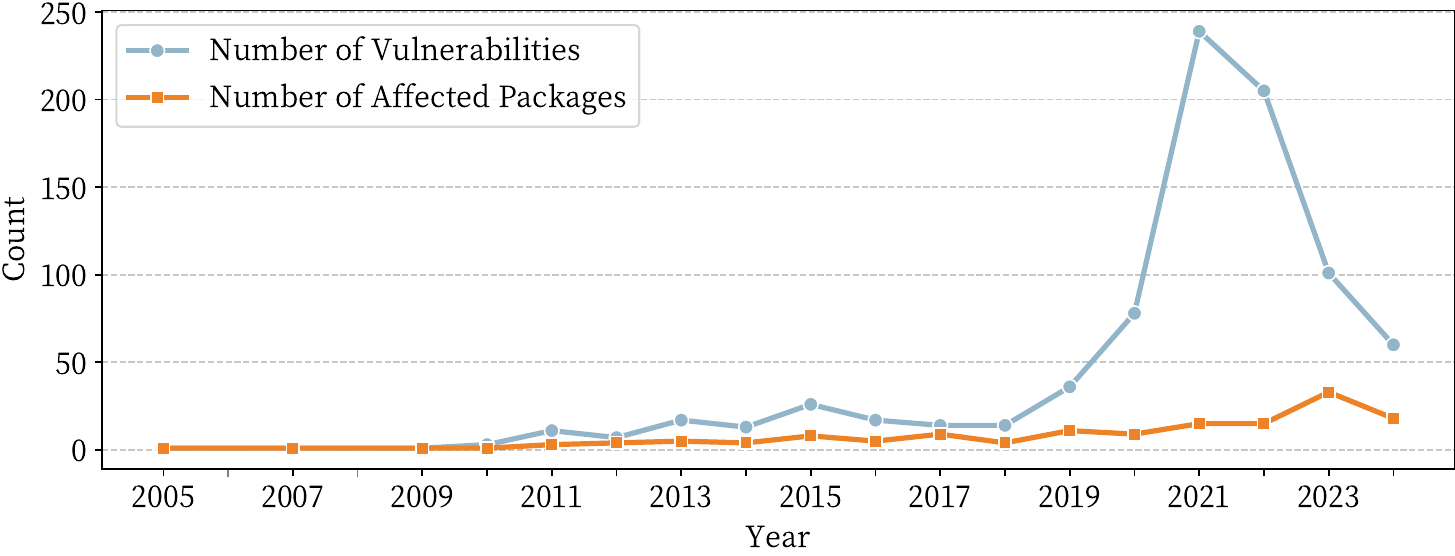}
  \caption{Trends in Vulnerabilities and Affected Packages in LLM Supply Chains}
    \label{fig:vulnerabilityTrends}
\end{figure}

As shown in Figure~\ref{fig:vulnerabilityTrends}, vulnerability reports were infrequent before 2019.\before{, with fewer than 20 reported annually.}
However, the number of vulnerabilities increased substantially after 2019, reaching 78 in 2020 and peaking at 239 in 2021.
\dx{This surge aligns with the rapid growth and deployment of LLMs, particularly following the introduction of the Transformer architecture~\cite{vaswani2017attention} in 2017 and the release of GPT-2~\cite{radford2019language} and GPT-3~\cite{brown2020language} between 2019 and 2020.}
\before{This sharp rise corresponds with the rapid development of LLMs, particularly following the release of GPT-3 by OpenAI in 2020, which marked a pivotal moment in the explosive growth of LLMs.}
A closer look at the 2021 data reveals that 84\% (200 of 239) of the vulnerabilities were attributed to a single core package, \texttt{TensorFlow}. This suggests that the peak in vulnerabilities was largely driven by issues within a few key dependencies, rather than a widespread increase across the supply chain. After 2021, the number of reported vulnerabilities began to decline. This decline can likely be attributed to the progressive resolution of vulnerabilities in widely used core dependencies, such as \texttt{TensorFlow}.
Despite the decline in reported vulnerabilities, the number of affected packages continued to rise. 
\dx{This can be attributed to the increasing complexity and expanding functionality of LLMs, which require diverse capabilities such as distributed training, efficient inference, API integration, and diagnostic logging. As discussed in Section~\ref{sec:RQ1}, numerous packages support system integration, networking, and diagnostics, thereby broadening the dependency surface. The growing reliance on external dependencies has expanded the risk landscape.}
This trend suggests that while individual vulnerabilities are being addressed, the overall security risk persists due to the growing interdependence of LLM ecosystems on external software.

\begin{figure}[h]
  \centering
  \includegraphics[width=\linewidth]{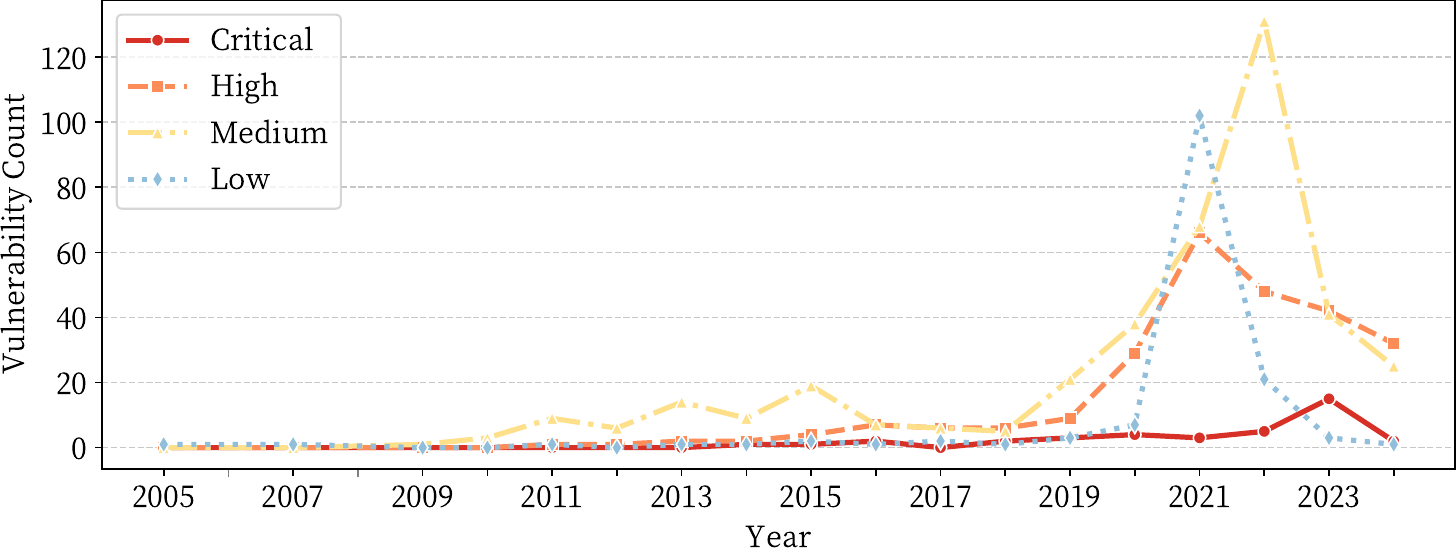}
  \caption{Vulnerability Severity Trends in LLM Supply Chains}
    \label{fig:vulnerabilitySeverityTrends}
\end{figure}

Figure~\ref{fig:vulnerabilitySeverityTrends} illustrates the annual distribution of LLM supply chain vulnerabilities by severity. Medium severity vulnerabilities constitute the largest proportion (47.75\%), followed by high severity vulnerabilities (30.21\%). Low severity vulnerabilities account for 17.54\%, while critical vulnerabilities represent a smaller but notable fraction (4.50\%). Vulnerabilities of all severities were relatively rare before 2015, with only sporadic occurrences of medium and low severity vulnerabilities. However, the number of vulnerabilities increased from 2015, particularly medium and high severity ones. The peak in 2021 saw a sharp rise in low severity vulnerabilities (102), driven primarily by a concentrated number of vulnerabilities in core packages (i.e., \texttt{TensorFlow}).

\begin{figure*}[h]
  \centering
  \includegraphics[width=.8\linewidth]{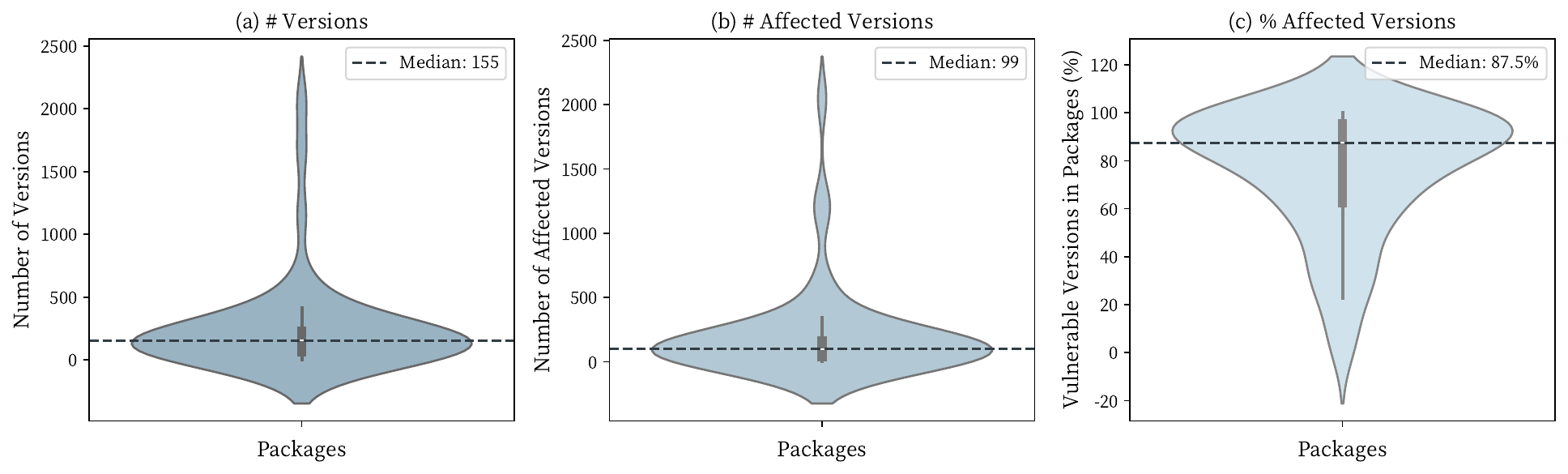}
  \caption{Distribution of Packages and Affected Packages of the Vulnerable Dependent Projects}
    \label{fig:affectedVersion}
\end{figure*}

Figure~\ref{fig:affectedVersion} illustrates the distribution of total versions (Fig.\ref{fig:affectedVersion}a), affected versions (Fig.\ref{fig:affectedVersion}b), and the percentage of affected versions (Fig.\ref{fig:affectedVersion}c) in vulnerable LLM supply chain packages.
Most packages have a large number of versions, with a median of 155 versions, of which 99 are affected by at least one vulnerability. Additionally, 87.5\% of package versions in vulnerable packages are impacted by vulnerabilities.
This finding highlights the pervasive nature of vulnerabilities across multiple versions of packages in the LLM supply chain. Such a widespread impact complicates the identification of unaffected versions for rollbacks in the absence of immediate fixes. Furthermore, the data suggest that packages with more versions tend to have a higher absolute number of affected versions, reinforcing the need for proactive vulnerability management and robust patching practices.

\begin{center}
    \resizebox{\linewidth}{!}{
\begin{tabular}{l!{\vrule width 1pt}p{0.9\columnwidth}}
    \makecell{{\LARGE \faLightbulbO}}  &\textbf{Summary.} \textit{Vulnerabilities in the LLM supply chain surges after 2015, peaking in 2021 due to vulnerabilities in core dependencies like TensorFlow, while the number of affected packages continues to rise. Most vulnerabilities are classified as medium or high severity, with a median of 87.5\% of package versions in vulnerable dependencies affected by at least one vulnerability.}
\end{tabular}}
\end{center}
\subsubsection{Vulnerability Type}
\label{sec:vulnerabilityType}
The LLM supply chain encompasses 121 distinct vulnerability types, with an average of 7.4 vulnerabilities per type. Table~\ref{tab:vulnerabilityType} presents the top ten vulnerability types and their severity distribution.

The data reveals that DoS vulnerabilities account for 262 instances, making it the most prevalent type, with the majority classified as medium severity (146). Their widespread occurrence poses substantial risks across varying severity levels and demands comprehensive mitigation efforts.
Other vulnerability types, such as Information Exposure (50 instances) and NULL Pointer Dereference (46 instances), highlight various dimensions of security risks, including the exposure of sensitive data and critical application failures.
Notably, while Arbitrary Code Execution vulnerabilities are limited to 29 instances, they exhibit a disproportionately high number of critical cases. These vulnerabilities enable attackers to remotely execute malicious code, potentially compromising entire systems or exposing sensitive data.
Overall, the LLM supply chain faces a multifaceted threat landscape. High-frequency vulnerabilities, such as DoS, require broad mitigation strategies, while those with concentrated severity (e.g., Out-of-Bounds and ReDoS) demand targeted, context-specific solutions.

\begin{table}[h!]
  \caption{Top 10 Vulnerability Types in LLM Supply Chain}
  \centering
  \label{tab:vulnerabilityType}
  \footnotesize
  \resizebox{\linewidth}{!}{
  \begin{tabular}{@{}l|cccc|c@{}}
    \toprule
    \textbf{Vulnerability Type} & \textbf{Low} & \textbf{Meduim} & \textbf{High} & \textbf{Critical} & \textbf{Total}  \\
    \midrule
    \textbf{DoS} & 69 & 146 & 46 & 1 & 262 \\
    \textbf{Out-of-Bounds} & 12 & 12 & 37 & 1 & 62 \\
    \textbf{Buffer Overflow} & 24 & 10 & 17 & 2 & 53 \\
    \textbf{Information Exposure} & 10 & 33 & 6 & 1 & 50  \\
    \textbf{NULL Pointer Dereference} & 9 & 18 & 19 & 0 & 46 \\
    \textbf{Improper Input Validation} & 5 & 21 & 15 & 0 & 41 \\
    \textbf{Cross-site Scripting (XSS)} & 1 & 28 & 1 & 0 & 30 \\
    \textbf{Arbitrary Code Execution} & 0 & 5 & 15 & 9 & 29 \\
    \textbf{Directory Traversal} & 4 & 11 & 13 & 1 & 29 \\
    \textbf{ReDoS} & 0 & 15 & 9 & 0 & 24 \\
    \bottomrule
  \end{tabular}}
\end{table}

We analyzed the distribution of vulnerability types across package categories and found that ``Machine Learning'' packages account for 91.1\% of all reported vulnerabilities, with a median of seven vulnerabilities per package.
``Machine Learning'' packages are predominantly affected by DoS vulnerabilities (206 cases) and memory-related issues, such as Out-of-Bounds (57 cases), Buffer Overflow (51 cases), and NULL Pointer Dereference (44 cases). In contrast, ``Network'' packages exhibit a broader range of vulnerabilities, including DoS (5 cases), Information Exposure (5 cases), and Directory Traversal (4 cases), underscoring challenges in securing data access and preventing unauthorized intrusions. ``Utility'' packages are mainly impacted by Regular Expression Denial of Service (ReDoS), with two cases, pointing to inefficiencies in regular expression handling. ``Data'' and ``Efficiency'' packages, while showing fewer vulnerabilities overall, tend to have vulnerabilities related to code execution, such as Arbitrary Code Execution and Command Injection, highlighting the risks associated with processing untrusted inputs.

\begin{center}
    \resizebox{\linewidth}{!}{
\begin{tabular}{l!{\vrule width 1pt}p{0.9\columnwidth}}
    \makecell{{\LARGE \faLightbulbO}}  &\textbf{Summary.} \textit{DoS is the most common vulnerability in the LLM supply chain. ``Machine Learning'' are the most vulnerable package type, comprising 91.1\% of all reported vulnerabilities.}\\
\end{tabular}}
\end{center}
\subsubsection{Vulnerability LifeCycle}
\label{sec:vulnerabilityLifeCycle}

\begin{figure}[h]
  \centering
  \includegraphics[width=\linewidth]{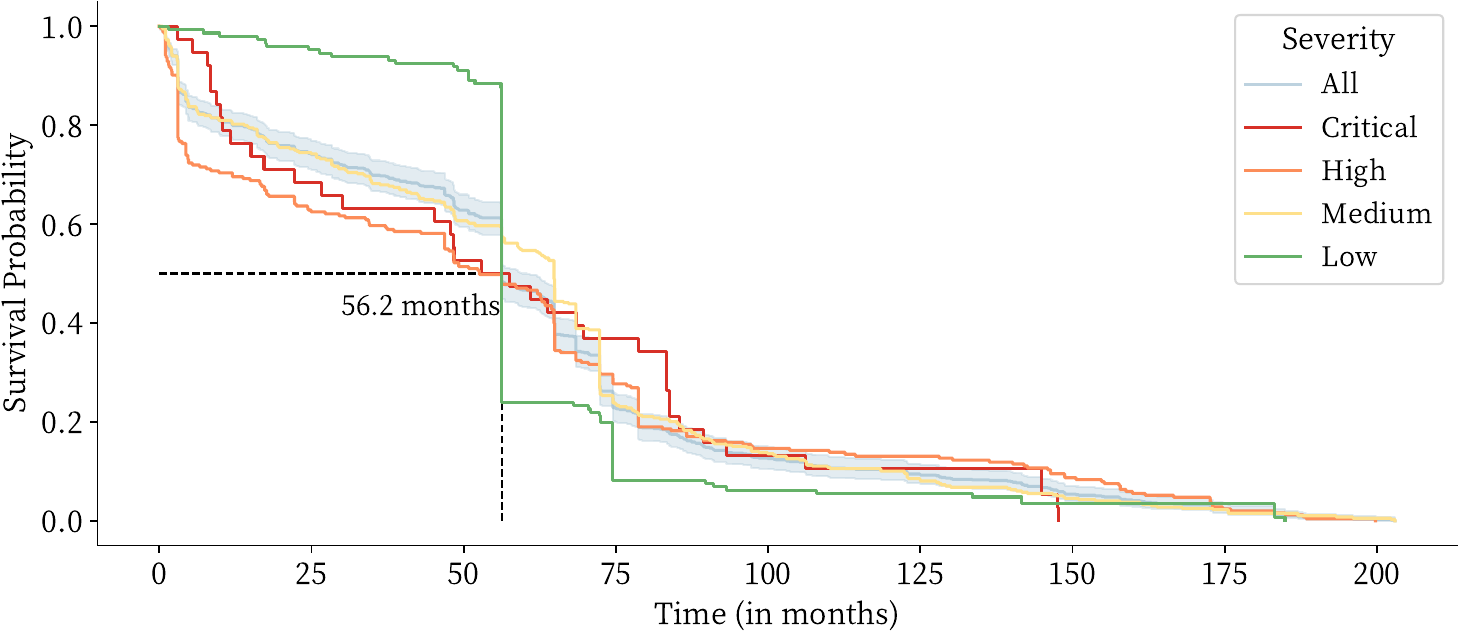}
  \caption{Survival Probability of Vulnerabilities in LLM Supply Chain from Introduction to Disclosure}
    \label{fig:affectToDisclose}
\end{figure}

Figure~\ref{fig:affectToDisclose} illustrates the vulnerability survival probability from the time of vulnerability introduction to disclosure. The sharp decline of all vulnerabilities in the initial months indicates that many vulnerabilities are disclosed shortly after their introduction. However, a long-tail distribution persists, with some vulnerabilities remaining undiscovered for over a decade. This prolonged survival time is likely due to difficulties in identifying vulnerabilities in legacy codebases, which often receive less scrutiny. The median time from introduction to disclosure is 56.2 months, highlighting long-term security risks within the LLM supply chain.
Critical vulnerabilities exhibit the shortest survival time, suggesting they are prioritized for discovery and disclosure due to their severe impact.
Interestingly, low severity vulnerabilities show a sharp decline in survival probability during the mid-period, primarily due to the release of \texttt{TensorFlow} version 2.1.4, which addressed 117 low severity vulnerabilities in a single update.

\begin{figure}[h]
  \centering
  \includegraphics[width=\linewidth]{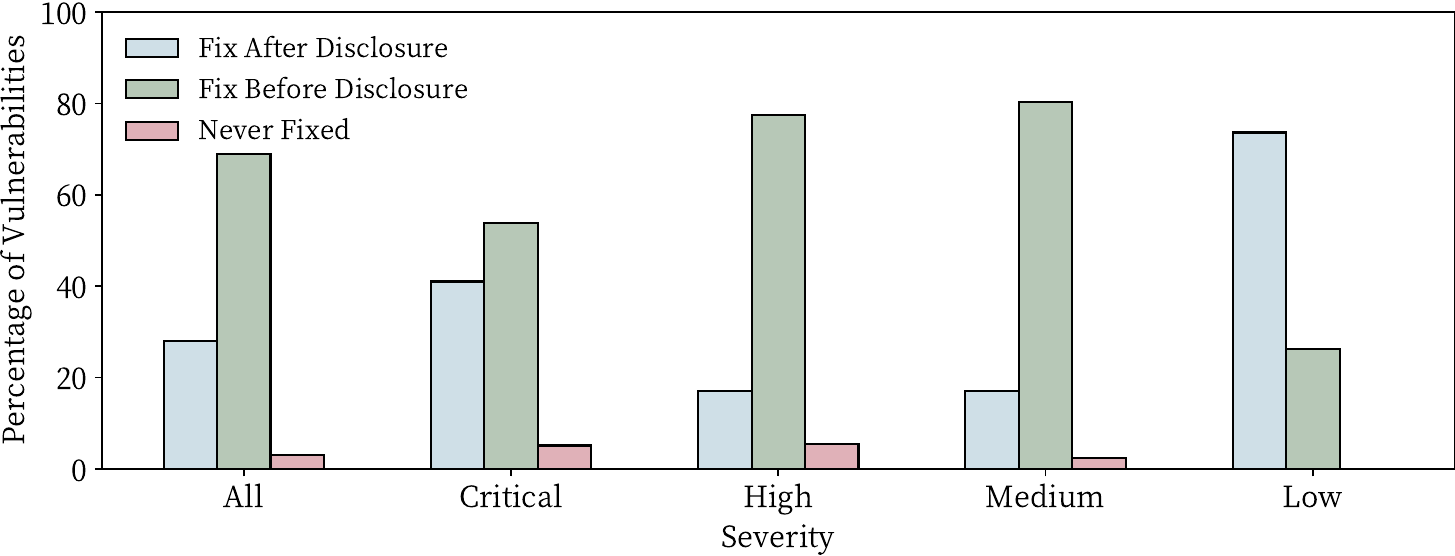}
  \caption{Proportional Distribution of Vulnerabilities by Severity from Disclosure to First Fix Release}
    \label{fig:fixStatusBySeverity}
\end{figure}

Figure~\ref{fig:fixStatusBySeverity} illustrates the fix status of vulnerabilities across different severity levels. Overall, 68.9\% of vulnerabilities are fixed before disclosure, while 28\% are fixed afterward, with a median fix time of 25 days. This delay provides attackers with a window of opportunity to exploit disclosed vulnerabilities in software packages. Critical vulnerabilities demonstrate the shortest median time to fix at 15 days, reflecting their prioritization due to their potential impact.
Additionally, compared to other severities, low severity vulnerabilities have a higher proportion (73.6\%) of ``Fix After Disclosure''.
\dx{To assess the statistical significance of this difference, we performed a chi-squared test~\cite{pearson1900x2} comparing the fix timing between low and non-low severity vulnerabilities. The result ($\chi^2$ = 174.91, df = 1, p $<$ 0.0001) indicates a significant association, suggesting that low-severity vulnerabilities are more likely to be fixed after public disclosure.}
This trend may result from their lower perceived security threat.
Furthermore, 3.1\% of vulnerabilities remain unfixed. 
\dx{To examine the relationship between fix status and project popularity, we compared GitHub star and fork counts for projects with fixed versus unfixed vulnerabilities. Prior to hypothesis testing, we applied Shapiro–Wilk tests~\cite{shapiro1965analysis} to assess the normality of the distributions. The results indicated significant deviations from normality in all cases (e.g., W = 0.8096, p $<$ 0.0001 for the star counts of ``Never Fixed'' projects), supporting the use of the non-parametric Mann–Whitney U test~\cite{mann1947test}.}
A Mann-Whitney U test~\cite{mann1947test} of community activity in projects associated with these vulnerabilities indicates that fixed vulnerabilities are linked to significantly higher median stars (63,144 vs. 13,079, U = 3626.0, p $<$ 0.0001) and forks (15,522 vs. 2,602, U = 4326.0, p $<$ 0.0001) compared to those that remain unfixed.

\begin{center}
    \resizebox{\linewidth}{!}{
\begin{tabular}{l!{\vrule width 1pt}p{0.9\columnwidth}}
    \makecell{{\LARGE \faLightbulbO}}  &\textbf{Summary.} \textit{Package vulnerabilities in the LLM supply chain take 4.7 years in the median to get disclosed. Additionally, 3.1\% of vulnerabilities remain unfixed and they are associated with less popular repositories. Low severity vulnerabilities are more likely to be fixed after disclosure.}\\
\end{tabular}}
\end{center}
\subsubsection{Influence on LLM}
\label{sec:influenceonLLM}
\dx{In this section, we explored the influence of vulnerabilities in the LLM supply chain on LLM from three perspectives: 
\ding{182} the proportion of surveyed LLMs that include vulnerable dependency versions,
\ding{183} the distribution of vulnerabilities within the LLM supply chain, and
\ding{184} the relationship between dependency usage frequency and vulnerability counts.}

Among the 52 LLMs analyzed, 33 included dependency configurations in the form of \texttt{requirements.txt} or \texttt{setup.py}. Upon examining these LLMs, we found that 23.2\% of the specified dependency version ranges were affected by known vulnerabilities.
For example, the strictly constrained dependencies of Starcoder, ``tqdm==4.65.0'' and ``transformers==4.28.1'', both had vulnerabilities.
In total, 75.8\% (25 out of 33) of LLMs with dependency configurations contained at least one vulnerable dependency version constraints. Specifically, 19.8\% of strict version constraints (i.e., ==) were found to be vulnerable, while 28.5\% of loose version constraints (i.e., those constrained within a range, such as \textgreater= or $<$) were affected. This highlights a critical trade-off in dependency management:
strict constraints enhance stability but do not eliminate vulnerabilities, whereas loose constraints increase flexibility but heighten security risks.
Among the affected dependencies, the \texttt{transformers} were the most vulnerable, with 82\% (18 out of 22) of its specified versions containing vulnerabilities.

The distribution of vulnerabilities in the LLM supply chain is highly imbalanced, with a small subset of packages accounting for most reported vulnerabilities.
Using the Gini coefficient~\cite{gini1912variability} as a measure of inequality, we observed a value of 0.81, indicating a significant imbalance in vulnerability distribution. Specifically, the top three packages—\texttt{TensorFlow}, \texttt{Django}, and \texttt{Ansible}—accounted for 71.01\% of all vulnerabilities (632 out of 890), despite comprising only 5.8\% of the total packages with vulnerabilities. In contrast, the remaining packages account for only 28.99\% of vulnerabilities, underscoring a ``long-tail'' distribution where the majority of packages are relatively less affected. This indicates that vulnerabilities are concentrated in a few critical packages, which likely serve as ``chokepoints'' in the software dependency network. The security of these high-vulnerability packages directly impacts numerous downstream systems.

We investigated whether dependency frequency correlates with vulnerability count. Our results show a Spearman correlation coefficient~\cite{spearman1961proof} of 0.1471 with a p-value of 0.2981, indicating a weak and statistically insignificant positive correlation between the frequency of dependency of a package and its vulnerability count.
\dx{To further examine potential differences, we categorized packages into two groups based on a quartile-derived threshold.} When comparing high-dependency packages (top 25\% of dependency frequency, frequency $>=$ 7) with low-dependency packages (bottom 25\%, frequency $<= 1$), we found that high-dependency packages had an average vulnerability count of 31.94, compared to just 11.62 for low-dependency packages. This suggests that high-dependency packages tend to be more vulnerable, even though the correlation is not statistically significant.
\begin{center}
    \resizebox{\linewidth}{!}{
\begin{tabular}{l!{\vrule width 1pt}p{0.9\columnwidth}}
    \makecell{{\LARGE \faLightbulbO}}  &\textbf{Summary.} \textit{75.8\% of LLMs with dependency configurations contain at least one vulnerable dependency in the dependency configuration file. Vulnerabilities in the LLM supply chain are highly concentrated, with a few critical packages accounting for 71\% of all vulnerabilities. Packages with high-frequency dependencies in LLMs are more likely to exhibit vulnerabilities.}\\
\end{tabular}}
\end{center}

\section{RQ3: Community Activity}
\label{sec:RQ3}
\dx{This section addresses the question: How are vulnerabilities in the LLM dependency supply chain addressed in practice?}

\subsection{Methodology}

In Section~\ref{sec:communityActivity}, we identified 100 merged PRs related to vulnerabilities across 52 LLMs. To focus on vulnerabilities specifically associated with the LLM dependency supply chain, we examined whether each PR was related to package version updates. Specifically, we analyzed the list of modified files in each PR. If the list included \texttt{requirements.txt} or \texttt{setup.py} and indicated updates to package versions, we classified the PR as related to package version updates.
For these vulnerability-related PRs associated with version updates, we further examined their vulnerability types. We compared the types of vulnerabilities discussed in these PRs with the prevalent types identified in Section~\ref{sec:vulnerabilityType} for the LLM dependency supply chain. This comparison enabled us to explore the gap between vulnerabilities addressed by developers and those in the dependency supply chain.
Additionally, we investigated the time difference between the PR merge date and the package version update date (i.e., the vulnerability fix date). To model this time difference, we applied the survival analysis method described in Section~\ref{sec:RQ2Method} to examine how long LLM maintainers take to update the package versions in their projects after the fixed package version updates.

\subsection{Results}
\subsubsection{Overview}
Among the 100 vulnerability-related PRs, there were 1,036 additions and 757 deletions.
There is a positive correlation between deletions and additions in merged PRs addressing vulnerabilities. In particular, 80\% of the PRs involve only an addition and a deletion.

Our dataset includes 52 LLM projects, but only 9 (17.3\%) contain vulnerability-related PRs. This indicates only a small subset of LLM projects actively addressing security issues. We further analyzed the relationship between the number of vulnerability-related PRs and the popularity of LLMs. The Pearson correlation coefficient~\cite{pearson1895note} between the number of PRs and stars of LLMs is 0.966 (p-value: 2.399e-05), and the correlation with forks is 0.985 (p-value: 1.313e-06), both statistically significant. These results suggest that more popular LLM projects are more likely to exhibit a higher number of vulnerability-related PRs.

\subsubsection{Version Update Vulnerabilities}

Among the 100 vulnerability-related PRs, 79 involved upgrading dependency versions. In terms of vulnerability types, the most frequent vulnerabilities were Out-of-Bounds (17 instances), Remote Code Execution (9 instances), Untrusted Code Inclusion (9 instances), and Cross-Site Scripting (9 instances). However, as shown in Table~\ref{tab:vulnerabilityType}, which presents types of vulnerabilities in the LLM supply chain, the vulnerabilities fixed through PR upgrades differ from the general distribution of vulnerabilities in the supply chain. Specifically, DoS, the most common vulnerability type in the supply chain, appears infrequently in PR fixes. In contrast, Remote Code Execution and Untrusted Code Inclusion, which are less common in the supply chain, are frequently addressed in PRs.
This suggests that LLM maintainers prioritize the fix of high-risk vulnerabilities, such as Out-of-Bounds and Remote Code Execution, as these directly impact the execution safety and functional stability of the model.

\begin{figure}[h]
  \centering
  \includegraphics[width=\linewidth]{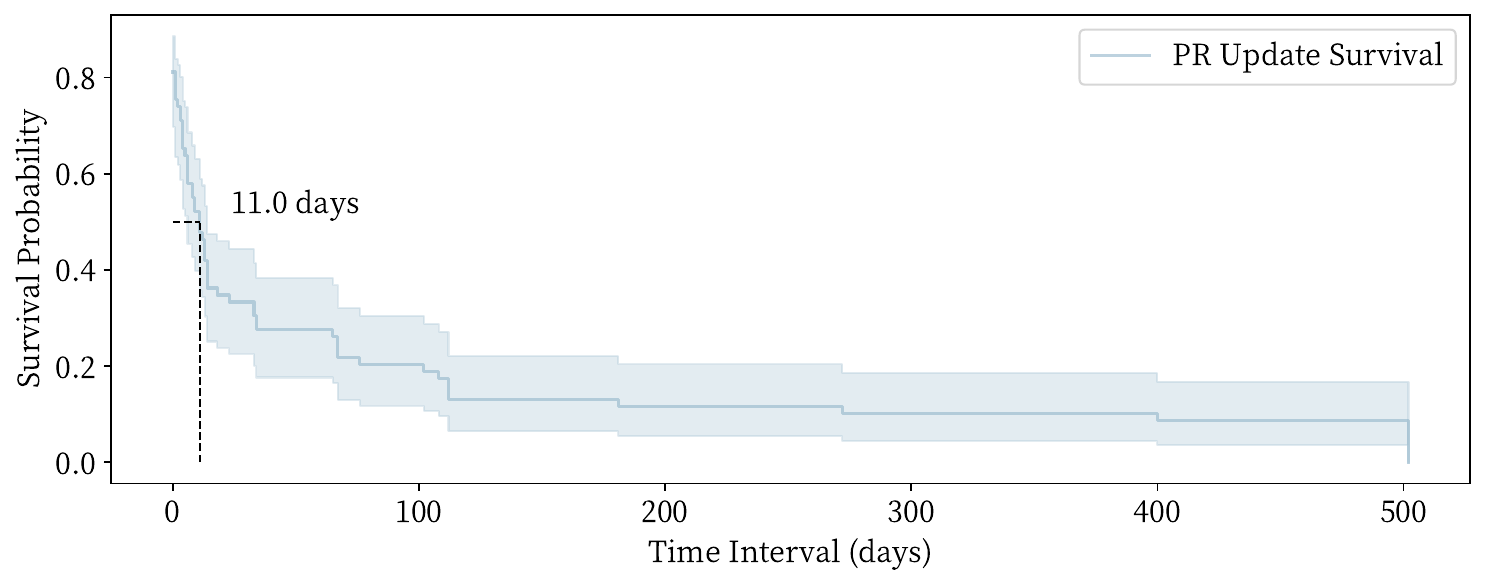}
  \caption{Survival Probability of Time Interval between Package Version Release and PR Updates}
    \label{fig:PRToVersionTime}
\end{figure}

The survival curve in Figure~\ref{fig:PRToVersionTime} illustrates the probability of PR updates over time after the release of a vulnerability-free version of a package in LLM projects. The rapid decline in survival probability suggests that most PR updates occur shortly after the release of the version. The curve flattens around 100 days, indicating that updates become less frequent over time. The maximum observed update interval is 502 days, reflecting significant delays in specific cases.
The median interval between version release and PR update is 11 days. As discussed in RQ~\ref{sec:vulnerabilityLifeCycle}, 28\% of the fixed versions of the packages are published after vulnerability disclosure, with a median time of 25 days. Additionally, new version releases often explicitly announce the vulnerabilities they fixed, potentially providing attackers with opportunities to exploit disclosed information. These two timelines, the disclosure-to-fix interval, and the fix-to-PR interval, together may allow attackers enough time to exploit known vulnerabilities, posing significant security risks to LLM projects.
\begin{center}
    \resizebox{\linewidth}{!}{
\begin{tabular}{l!{\vrule width 1pt}p{0.9\columnwidth}}
    \makecell{{\LARGE \faLightbulbO}}  &\textbf{Summary.} \textit{Vulnerabilities in the LLM supply chain are a significant concern, with 79\% of vulnerability-related PRs associated with dependency version updates. These updates occur within a median of 11 days after the release of a fixed version.}\\
\end{tabular}}
\end{center}

\section{Discussion and Implication}
\label{sec:discussionAndImplication}
In this section, we compare vulnerabilities in the LLM dependency supply chain with those in the Python ecosystem (Section~\ref{sec:comparison}). Then, we highlight the implications of our study to researchers and practitioners (Section~\ref{sec:implication}).

\subsection{Comparison to the Python Ecosystem}
\label{sec:comparison}
In Section~\ref{sec:RQ2}, we examined vulnerabilities within the LLM dependency supply chain. However, how do these vulnerabilities differ from those in the broader PyPI ecosystem? To identify the unique characteristics of vulnerabilities in the LLM dependency supply chain, we conducted a comparative analysis. Specifically, we compared our findings with the empirical study by Alfadel et al.~\cite{alfadel2023empirical}, which analyzed 1,396 vulnerability reports across 698 Python packages within the PyPI ecosystem. This comparison provides valuable insights into the distinct features and risks associated with the LLM dependency supply chain, highlighting its unique challenges relative to the broader Python ecosystem.

\subsubsection{Vulnerability Trends}
In terms of vulnerability temporal trends, both the number of vulnerabilities and the number of affected packages have been increasing over time in the Python ecosystem. In contrast, the LLM ecosystem saw a peak in vulnerability reports in 2021, primarily driven by a surge in vulnerabilities associated with a few key dependencies, such as \texttt{TensorFlow}. Although the number of vulnerabilities in the LLM ecosystem has declined since then, the number of affected packages has continued to increase steadily.
Furthermore, vulnerabilities in the LLM ecosystem are highly concentrated in a small number of dependencies, whereas those in the Python ecosystem are more evenly distributed across packages. This highlights the disproportionate impact of individual core dependencies on the LLM ecosystem.

We observed that most vulnerabilities in both the LLM and Python ecosystems fell into the medium or high severity categories. However, the LLM ecosystem exhibited a lower proportion of medium severity vulnerabilities (47.75\% vs. 64.03\%) but a higher proportion of high severity vulnerabilities (30.21\% vs. 25.42\%). This trend indicates that vulnerabilities in the LLM ecosystem are generally more severe, posing a potentially greater security risk compared to the Python ecosystem. 
\dx{A notable case is CVE‑2025‑32434, a critical RCE vulnerability in PyTorch affecting versions $\leq$ 2.5.1 via \texttt{torch.load}. Despite being previously thought safe, this deserialization path allowed arbitrary code execution from malicious model weight files. Since LLM workflows commonly load external checkpoints, this vulnerability potentially enables attackers to execute arbitrary code, exfiltrate data, or embed backdoors undetected.}

Both the LLM and Python ecosystems showed extensive vulnerability exposure, with about half of the packages in each ecosystem having at least 80\% of their versions affected. However, the LLM ecosystem had a higher median number of versions per package (155 vs. 40), reflecting greater versioning complexity and increased challenges in vulnerability management.

\subsubsection{Vulnerability Type}
Our analysis revealed differences in the prevalence of vulnerability types between the Python and LLM ecosystems. While XSS was the most common vulnerability in Python, it ranked only seventh in the LLM ecosystem. This disparity might arise from the widespread use of Python packages in web development and interactive applications, where XSS is more common.
\dx{In contrast, the LLM ecosystem primarily focuses on machine learning tasks.  However, some LLM repositories incorporate web-facing components for model deployment or visualization (e.g., \texttt{Gradio}, \texttt{Streamlit}, and \texttt{Aiohttp}) where XSS vulnerabilities remain relevant. These packages are commonly used to provide interactive interfaces for inference or training monitoring, potentially exposing attack surfaces if not properly secured.}
DoS vulnerabilities were prominent in both ecosystems, ranking first in the LLM ecosystem and second in Python, underscoring their critical impact on the functionality of both machine-learning and general-purpose Python applications.
\before{In contrast, the LLM ecosystem, centered on machine learning, faced fewer vulnerabilities related to user input handling.}

\subsubsection{Vulnerability Lifecycle}
Vulnerabilities in the Python ecosystem were discovered more quickly, with a median time of 39 months compared to 56.2 months in the LLM ecosystem. This discrepancy might since dependencies in the LLM ecosystem often lacked the same level of community scrutiny and reporting mechanisms found in more mature Python packages. The extended discovery period increased security risks, allowing attackers a larger window for exploitation. Additionally, both ecosystems exhibited a long-tail effect in the vulnerability discovery timeline, with some vulnerabilities remaining undetected for over 180 months. This highlights the need for more proactive detection mechanisms to shorten discovery times and mitigate long-term security risks.

\subsection{Implications}
\label{sec:implication}
\dx{To support practitioners in managing dependencies more securely and effectively, we propose the following guidelines based on our empirical findings:}

\noindent\dx{\textbf{Regularly audit and monitor vulnerabilities in dependencies with high frequency and concentrated security issues.}}
Vulnerabilities are concentrated in a small subset of packages within the LLM dependency supply chain. For instance, \texttt{TensorFlow}, \texttt{Django}, and \texttt{Ansible} account for 71.01\% of all vulnerabilities in the LLM ecosystem. Developers should prioritize these high-risk dependencies due to their widespread use and the significant impact of their vulnerabilities. As noted in Section~\ref{sec:RQ1}, only 11\% of packages are used more than 10 times, while 60.1\% are used only once. Vulnerabilities in frequently used packages can have a broader impact on the LLM ecosystem, making it crucial for the community to focus on securing these high-frequency dependencies. 

\noindent\dx{\textbf{Accelerate vulnerability discovery through automation.}}
As discussed in Section~\ref{sec:comparison}, the median time to discover vulnerabilities in the LLM ecosystem (56.2 months) is longer than in the Python ecosystem (39 months). This prolonged discovery period increases the window of opportunity for attackers to exploit vulnerabilities. To mitigate this risk, the LLM community should prioritize enhancing the speed and effectiveness of vulnerability detection. Implementing automated vulnerability scanning tools and real-time reporting mechanisms could help shorten discovery times~\cite{lin2024vulnerabilities}.

\noindent\textbf{Minimize the use of vulnerable dependency versions.}
As highlighted in Section~\ref{sec:vulnerabilityTrends}, the median proportion of affected versions in packages with reported vulnerabilities is 87.5\%. Most of the package versions are affected by the vulnerability. Users should carefully select dependency packages and avoid versions with known vulnerabilities. Package maintainers need to clearly label vulnerable versions and provide explicit guidance to users.
Additionally, developers should regularly check for updates and patches to ensure they are using the most secure versions of their dependencies.

\noindent\textbf{Exercise caution when using less popular packages with unfixed vulnerabilities.}
As noted in Section~\ref{sec:vulnerabilityLifeCycle}, vulnerabilities that remain unfixed are often associated with less popular packages which typically receive less attention from maintainers. Developers should be careful when incorporating such packages into their projects, as they may expose the system to unfixed vulnerabilities.
When selecting dependencies, developers should prioritize packages with active maintainers and a history of timely updates. If using less popular packages is unavoidable, developers should closely monitor security advisories and be prepared to mitigate potential risks.

\noindent\textbf{Strengthen dependency management practices.}
Our findings indicate that 75.8\% of LLMs with dependency configurations contain at least one vulnerable dependency version (Section~\ref{sec:influenceonLLM}).
To manage dependencies effectively, developers can use tools such as Dependabot~\cite{dependabot} or Snyk~\cite{snyk_security}, which automate the tracking and updating of dependencies. Additionally, maintainers should regularly review their dependency configurations, identify potential vulnerabilities, and apply timely updates to mitigate security risks.

\section{Threats to Validity}
\label{sec:threats}

Internal validity concerns the threats to how we perform our study. One potential threat stems from the assumption that identified vulnerabilities in LLM dependencies directly affect the corresponding LLM projects. LLM may include a dependency but not necessarily invoke the vulnerable functionality, making it challenging to assess the actual security risk.
\dx{However, we mitigate this threat by focusing on potential vulnerability exposure rather than confirmed exploitability. Prior studies such as Decan et al.~\cite{decan2018impact} and Alfadel et al.~\cite{alfadel2023empirical} also adopt this assumption, acknowledging that dependency-level vulnerabilities can propagate risks downstream. Moreover, LLMs are frequently deployed as components of broader systems (e.g., involving preprocessing, model serving, and user interfaces) that may indirectly invoke vulnerable functionality. Given the limited detail in many vulnerability disclosures regarding exploit conditions, we consider this approach to be a conservative yet practical approximation. Future work could further reduce this threat by employing dynamic analysis or code-level reachability techniques to validate whether and how vulnerabilities are triggered in real-world scenarios.}
Another concern is the use of the Kaplan-Meier estimator for time-to-event analysis. This method assumes that missing or incomplete data are random and independent; if this assumption is violated, the results may be biased. Additionally, it does not account for confounding variables or group differences, which could influence the accuracy of our findings. \dx{However, our study does not rely on causal inference from the survival curves. Instead, we use Kaplan–Meier analysis to characterize general trends (e.g., the presence of long-tail delays in vulnerability disclosure) that remain robust even under minor violations of the random censoring assumption. This approach aligns with prior studies~\cite{decan2018impact,alfadel2023empirical}, which employed Kaplan–Meier estimators for exploratory analyses of vulnerability timelines in software ecosystems.}
\before{However, the mere presence of these vulnerabilities warrants attention. LLMs are frequently integrated into broader systems that include pre-processing, post-processing, and web-serving components, potentially reintroducing exploit paths for these vulnerabilities.
Moreover, many vulnerability reports did not provide clear details on how exactly a vulnerability is triggered, making it difficult to determine if a specific vulnerability posed a real risk to the LLM. Consequently, our study presented potential vulnerability risks in the analyzed LLM projects. Future research could benefit from incorporating runtime analysis or empirical testing to assess whether and how these vulnerabilities are triggered in practice.}

External validity concerns the threats to generalize our findings. A key limitation is the completeness of our dataset, as we primarily relied on Hou et al.~\cite{hou2023large}, GitHub searches, and Google queries to collect open-source LLMs. This approach may have overlooked lesser-known or non-public repositories, particularly those not widely cited in academic literature or hosted outside GitHub.
\dx{However, this focus aligns with our objective of analyzing real-world LLM dependency supply chain, which requires access to the source code and related community activities of models. Unlike other platforms such as Hugging Face, GitHub provides the complete source code and development history of LLMs. Furthermore, as described in Section~\ref{sec:datacollection}, we employed a multi-pronged search strategy to ensure comprehensive coverage of major open-source LLM projects. While our dataset may not be exhaustive, it is representative of the most active and influential projects in the current LLM ecosystem.}
Additionally, our study relies on vulnerability severity classifications from Snyk.io, which may differ from other sources, such as PyPI security advisories. However, since PyPI’s severity data is not publicly available, we relied solely on Snyk.io, a dataset frequently used in prior research~\cite{alfadel2023empirical,8530065,Lagsintherelease}.

\section{Related Work}
\label{sec:relatedwork}
\subsection{LLM Security}
There have already been several LLM surveys (e.g., ~\cite{gupta2023chatgpt,derner2023security,shayegani2023survey,dash2023chatgpt,derner2023beyond,renaud2023chatgpt,chen2024security,huang2024lifting,yang2024robustness}) related to security. \dx{These surveys have broadly examined the security threats faced by LLMs, covering topics such as adversarial attacks, data leakage, misuse risks, and mitigation strategies.}

Yao et al.~\cite{yao2024survey} surveyed the security implications of LLMs, analyzing their benefits, risks, vulnerabilities, and defense mechanisms. They concluded that LLMs contribute more positively than negatively to security, with user-level attacks being the most prevalent due to LLMs’ human-like reasoning. Similarly, Dong et al.~\cite{dong2024attacks} examined conversational LLM safety, identifying adversarial prompts and backdoor attacks as major threats at both inference and training stages. While their work reviews existing literature on LLM attacks and defenses, our study empirically analyzes real-world vulnerabilities in the LLM dependency supply chain.

Wang et al.~\cite{wang2024large} provided a comprehensive overview of the LLM supply chain, identifying its three key components: infrastructure, model lifecycle, and downstream application ecosystem. Their work emphasized the challenges and opportunities associated with ensuring the robustness, security, and ethical deployment of LLM systems across all stages. However, their analysis remains high-level, focusing on conceptual challenges without exploring specific vulnerabilities within third-party libraries or dependencies. Hu et al.~\cite{hu2024large} examined security risks within the LLM supply chain, identifying 12 key risks, including vulnerabilities in frameworks and third-party libraries. While their study provides a valuable discussion of potential attack vectors, it lacks empirical insights into real-world vulnerabilities and their impacts. Our research addresses these gaps by conducting an empirical analysis of package dependencies in the LLM supply chain, focusing on identifying and categorizing vulnerabilities in third-party libraries to provide actionable insights to secure LLM ecosystems.

\subsection{Dependency Supply Chain}
There are many studies that focus on vulnerabilities or bugs in the dependency supply chain.
Tan et al.~\cite{tan2022exploratory} conducted an exploratory study on the supply chains of deep learning frameworks, focusing on TensorFlow and PyTorch. They highlighted the hierarchical and sparse nature of these supply chains, which comprise dependent packages and downstream projects. Their analysis identified vulnerabilities within the ecosystem, revealing that a significant number of projects rely on a small set of critical nodes. In contrast, our study focuses on the specific security risks posed by third-party libraries within the supply chain rather than on structural vulnerabilities.
Alfadel et al.~\cite{alfadel2023empirical} analyzed security vulnerabilities in Python packages hosted on PyPI, examining the vulnerability lifecycle from discovery to resolution. Their study revealed that vulnerabilities often remain unidentified for years, with 40.86\% only being fixed after public disclosure, leaving dependent projects exposed to prolonged risks. Building on these findings, our research focuses on specific vulnerabilities within the package dependencies of LLMs.
Jiang et al.~\cite{jiang2022empirical} examined the supply chain of pre-trained models (PTMs), analyzing artifacts and security risks across eight model hubs (e.g., Hugging Face and TensorFlow Hub). Their study highlighted the hierarchical structure of PTM ecosystems, emphasizing discrepancies in model documentation and the role of maintainers as key risk factors. However, their focus was on dependencies between PTMs and datasets, overlooking the role of third-party libraries. In contrast, our study investigates Python package dependencies and their impact on the LLM supply chain.

\section{Conclusion}
\label{sec:conclusion}
The security of the LLM dependency supply chain is a critical aspect of overall LLM security. In this study, we empirically analyzed the vulnerabilities within the LLM dependency supply chain by examining 52 open-source LLMs and their third-party dependencies. Our findings reveal that half of the vulnerabilities remain undisclosed for over 56.2 months, significantly longer than the 39 months observed in the Python ecosystem. Additionally, 75.8\% of LLMs include at least one vulnerable dependency in the dependency configuration. Notably, machine learning packages present the greatest risk, with DoS vulnerabilities being the most common (32.5\%).
We also examine activities within LLM repositories to understand how maintainers address vulnerabilities in practice. Our analysis shows that proactive vulnerability fixes are rare: only 17.3\% of LLMs have PRs related to vulnerabilities, with 79\% of fixes achievable through dependency updates.
We recommend that practitioners focus on packages with concentrated vulnerabilities and high dependency frequencies, accelerate vulnerability discovery within the LLM ecosystem, minimize the use of vulnerable dependency versions, and strengthen dependency management practices. In conclusion, this study underscores the critical importance of securing the LLM dependency supply chain. Based on this study, the LLM community can mitigate threats in the LLM dependency supply chain and promote a more reliable LLM ecosystem.

\dx{Future work could explore the actual exploitability of dependency supply chain vulnerabilities in LLMs through dynamic or runtime analysis. Additionally, research may focus on developing automated dependency recommendation systems tailored to the unique needs of LLM projects. Another promising direction is the design of proactive monitoring frameworks that alert developers to newly disclosed vulnerabilities relevant to their specific dependency configurations.}

\bibliographystyle{ACM-Reference-Format}
\bibliography{references} 

\end{document}